\documentclass[%
 reprint,
 nofootinbib,
 amsmath,amssymb,
 aps,
 prd,
]{revtex4-2}
\usepackage{placeins}
\usepackage{graphicx}
\usepackage{dcolumn}
\usepackage{bm}
\usepackage{hyperref}
\usepackage{mathtools}

\usepackage{xcolor}
\usepackage{soul}

\usepackage{multirow}
\usepackage{lipsum}
\usepackage{makecell}
\usepackage{adjustbox}
\usepackage{resizegather}
\usepackage{rotating}

\newcommand{\dd}{\mathrm{d}}
\newcommand{\bara}{\bar a}
\newcommand{\barN}{\bar N}
\newtheorem{proposition}{Proposition}
\newtheorem{remark}{Remark}
\begin{document}
\title{{\bf Can the Universe Change Signature When Gravity is Dynamical?}}


\author{Yaghoub Heydarzade}
 \email{yheydarzade@bilkent.edu.tr}

\affiliation{
Department of Mathematics, Faculty of Sciences
Bilkent University, 06800 Ankara, Turkey
}

\begin{abstract}
Can a universe undergo a regular Euclidean--Lorentzian signature transition when the gravitational coupling itself is dynamical? We address this question in scalar--tensor gravity with nonminimal coupling \(F(\phi)R\). Although the spatially flat  Friedmann–Lemaître–Robertson–Walker (FLRW) sector can be mapped to the Einstein frame for \(F>0\) and a nondegenerate scalar redefinition, not all transition properties are frame independent. For a finite, positive, and sufficiently regular conformal factor, the existence and transverse character of the type change are preserved, whereas the extrinsic geometry is not. We therefore use the Einstein frame as a solution-generating representation and impose total geodesy in the physical Jordan frame.
We construct two exactly integrable classes of solutions. In the oscillator--ghost-oscillator branch, the scalar field is stationary at the transition and Jordan-frame total geodesy follows under suitable regularity assumptions on the conformal factor. In the critical exponential-potential branch, the scalar is generically nonstationary and total geodesy instead requires
\[
H_E\big|_\Sigma
=
\frac12
\left(
\frac{d\ln F}{d\psi}
\right)_\Sigma
\dot\psi_\Sigma .
\]
We also distinguish this condition from the stronger smoothness requirements of a Kossowski--Kriele-type transverse metric. Finally, while the canonical Einstein-frame scalar does not allow effective phantom evolution, the dynamical nonminimal coupling can generate a locally superaccelerating Jordan-frame regime near the transition when $
\left(\frac{d\ln F}{d\psi}\right)_\Sigma>0
$
for the chosen oscillator branch. Thus signature change persists beyond Einstein gravity, with regularity and effective cosmological behavior remaining intrinsically frame sensitive.
\end{abstract}

\maketitle
\vspace{0.5cm}

\section{Introduction}

The standard formulation of general relativity assumes that spacetime is a smooth four-dimensional manifold equipped with a Lorentzian metric. The Lorentzian signature fixes the causal structure of spacetime and is responsible for the hyperbolic character of the gravitational field equations. Nevertheless, both quantum-cosmological and classical considerations suggest that Lorentzian signature need not be fundamental at all stages of cosmic evolution. In the Hartle--Hawking no-boundary proposal, for example, the early universe is associated with a compact Euclidean regime from which a Lorentzian universe may emerge \cite{HartleHawking}. Related ideas on cosmological transitions involving a change of metric signature were also discussed in early work by Sakharov \cite{Sakharov}. More generally, signature-changing geometries provide a classical setting in which Euclidean and Lorentzian regions are joined across a hypersurface where the metric becomes degenerate \cite{EllisEtAl,Hayward,DrayEllisHellabyManogue,WhiteMagueijoVisser}.

Classical signature change in general relativity has been studied by allowing the metric to undergo a change of type across a transition hypersurface. Since the determinant of the metric vanishes on this hypersurface, the usual nondegenerate formulation of the field equations and the standard junction conditions cannot be applied naively. A geometrically precise framework was developed by Kossowski and Kriele, who formulated regularity conditions for transverse type-changing metrics \cite{KossowskiKriele}. In homogeneous and isotropic cosmology, these conditions imply that the transition hypersurface must be totally geodesic. Equivalently, the extrinsic curvature of the limiting nondegenerate hypersurfaces must vanish at the signature-changing surface. It should be emphasized, however, that different approaches to classical signature change exist, and the precise junction conditions can depend on how the field equations are extended to the degenerate hypersurface \cite{HellabyDray1995,DrayHellaby,HartleyTuckerTuckeyDray}.

A particularly useful analytic realization was given by Dereli and Tucker \cite{DereliTucker}. They showed that Einstein gravity coupled to a self-interacting scalar field admits exact signature-changing FLRW solutions. The central observation is that the reduced Einstein--scalar minisuperspace Lagrangian can be transformed, through a hyperbolic change of variables, into an oscillator--ghost-oscillator mechanical system. The lapse variation imposes the Hamiltonian constraint, selecting the zero-energy sector of the mechanical system. This construction yields exact solutions for the minisuperspace variables and hence for the scale factor and scalar field across a Euclidean--Lorentzian transition. Related work has also explored scalar fields, boundary conditions, particle production, and quantum-cosmological aspects of signature change \cite{DrayManogueTucker1991,DrayManogueTucker1993,DrayManogueTucker1995,DereliOnderTucker}.

A complementary line of investigation treats the spacetime signature
itself as a dynamical degree of freedom. In the effective-action
formulation initiated by Percacci and subsequently developed by
Greensite \cite{Percacci1991,Greensite1993}, the signature is
parametrized by a dynamical Wick angle whose preferred value is
determined from a quantum effective potential. Elizalde, Odintsov and
Romeo \cite{Elizalde1994}, and later Odintsov, Romeo and Tucker
\cite{Odintsov1994}, showed that this dynamical selection can be
sensitive to the topology of the background. In particular, for
compactified spacetimes of the form
$\mathbb{R}^{D-1}\times S^{1}$, the compactification scale enters the
one-loop effective potential explicitly, so that the preferred
signature may depend on the compactification radius, the quantum field
content, and the spacetime dimension.
\\
The setting considered in the present work is different. We restrict
attention to a classical four-dimensional FLRW minisuperspace with
fixed spatial topology, without a dynamical compactification modulus or
quantum effective potential. Within the spatially flat homogeneous
truncation used below, replacing the noncompact flat spatial sections
by a fixed compact flat quotient, while retaining the same local
$k=0$ geometry and homogeneous degrees of freedom, changes the reduced
action only by an overall constant comoving-volume factor. It therefore
does not modify the local minisuperspace equations or the
signature-change conditions studied here. Topology-dependent quantum
effects or dynamical compactification consequently lie beyond the
scope of the present classical analysis. It is nevertheless worth
noting that Ref.~\cite{Odintsov1994} explicitly remarks that the quantum
preference for a given signature may be modified once gravitational
effects are included.

 Returning to the classical minisuperspace setting, the mechanism
introduced by Dereli and Tucker was recently extended to a trace-coupled
modified gravity model of the form \(f(R,T_\phi)\), where \(T_\phi\) is
the trace of the scalar-field energy-momentum tensor
\cite{HazinedarHeydarzade}. In that model, the matter--geometry coupling modifies the scalar kinetic and potential sectors of the minisuperspace Lagrangian while preserving the possibility of an integrable hyperbolic representation. Regular signature-changing branches were shown to survive in the presence of trace coupling. Moreover, the \(f(R,T_\phi)\) model contains additional exact sectors beyond the original Dereli--Tucker oscillator branch, including a critical exponential-potential branch.

The purpose of the present paper is to ask whether this construction survives in a genuinely geometrical modified theory of gravity. The natural first candidate is scalar--tensor gravity. Scalar--tensor theories generalize general relativity by replacing the constant gravitational coupling with a scalar-dependent nonminimal coupling, and include the Brans--Dicke theory as their prototype \cite{BransDicke,FujiiMaeda,FaraoniBook}. Unlike the trace-coupled \(f(R,T_\phi)\) model, where the modification enters through the matter trace, scalar--tensor gravity modifies the gravitational sector itself by replacing the Einstein--Hilbert coupling with a nonminimal coupling \(F(\phi)R\). The effective Newton coupling is therefore dynamical, with
\begin{equation}
G_{\rm eff}\sim F(\phi)^{-1}.
\end{equation}
The question posed in the title can then be stated more precisely: can a Kossowski--Kriele regular Euclidean--Lorentzian transition occur when the gravitational coupling itself is a dynamical field?

This question is nontrivial because not all properties of a
signature-changing transition are conformally invariant. For a finite,
positive, and sufficiently regular conformal factor, the existence,
location, and transverse character of the type change are preserved,
whereas the extrinsic geometry of the transition hypersurface, and in
particular the total-geodesy condition, is generally frame dependent.
When \(F(\phi)>0\) and the scalar kinetic sector is nondegenerate,
scalar--tensor gravity can be mapped to the Einstein frame, where the spatially flat FLRW minisuperspace dynamics takes the standard Einstein--scalar form. However, this does not make the problem equivalent to minimally coupled Einstein gravity. The physical interpretation of the Jordan and Einstein frames in scalar--tensor gravity has been widely discussed \cite{Faraoni1999,FaraoniNadeau}. In the present problem, the issue is sharper because signature change occurs precisely at a hypersurface where the metric is degenerate, and the behavior of the conformal factor at this hypersurface affects the regularity of the physical geometry. Throughout this work we regard the Jordan frame as the physical frame. The Einstein frame is used as a solution-generating representation, while the final signature-change and total-geodesy conditions are imposed on the Jordan-frame metric.

A second technical point is also essential. The minisuperspace equations are most conveniently solved in a nondegenerate gauge time, denoted by \(\tau\), which may be chosen by setting the Einstein-frame lapse to unity. The signature-changing geometry, however, is naturally described by a coordinate \(\beta\) for which the lapse vanishes at the transition. For example, in the Einstein frame one may write
\begin{equation}
d\bar{s}^{2}
=
-\beta\,d\beta^{2}
+
\bar a(\beta)^{2}d\vec{x}^{2}.
\end{equation}
For \(\beta>0\), this metric has Lorentzian signature, whereas for \(\beta<0\) it has Euclidean signature, with the adopted sign convention. On the Lorentzian side one has
\begin{equation}
d\tau=\sqrt{\beta}\,d\beta,
\qquad
\tau-\tau_\Sigma=\frac{2}{3}\beta^{3/2}.
\end{equation}
Across the transition this relation must be specified as a real pullback prescription, not merely as a formal analytic continuation. We shall use a real transverse coordinate for which \(\tau-\tau_\Sigma=\frac23\operatorname{sgn}(\beta)|\beta|^{3/2}\), unless explicitly stated otherwise. This makes the lapse vanish linearly in \(\beta\), but it also shows why nonstationary scalar branches require an additional smoothness check: functions that are linear in \(\tau-\tau_\Sigma\) become only fractional-power functions of \(\beta\). Thus the solutions found in the gauge \(\bar N=1\) must first be treated as functions of \(\tau\), and only afterward pulled back to the signature-changing coordinate \(\beta\).

We construct two classes of exact scalar--tensor signature-changing cosmologies. The first is an oscillator branch, directly generalizing the Dereli--Tucker construction. In this branch the Einstein-frame scalar field is stationary at the transition. Consequently, if the conformal factor is finite and positive, the Jordan-frame transition remains regular and the transition hypersurface is totally geodesic. The second class is a critical exponential-potential branch. Exponential potentials are standard in scalar-field cosmology and appear in many inflationary and scaling-solution contexts \cite{LucchinMatarrese,RatraPeebles,CopelandLiddleWands}. Their role here is more specific: at the critical slope the light-cone minisuperspace system becomes exactly integrable. This branch is qualitatively different from the oscillator branch because the scalar field is not stationary at the transition. As a result, Jordan-frame total geodesy imposes a genuinely scalar--tensor condition involving the derivative of \(\ln F\) with respect to the Einstein-frame scalar.

The main result may be summarized as follows. Classical signature change can persist when the gravitational coupling is dynamical, but its regularity is not simply inherited from the Einstein frame. In scalar--tensor gravity, the conformal factor is part of the physical problem: it can preserve the regular transition in the stationary oscillator branch, while in the nonstationary exponential branch it imposes a new coupling-dependent regularity condition.

The paper is organized as follows. In Sec.~\ref{sec:STsetup} we introduce the scalar--tensor action and derive the spatially flat FLRW minisuperspace Lagrangian. In Sec.~\ref{sec:EinsteinFrame} we transform the system to the Einstein frame and identify the Einstein--scalar minisuperspace structure. In Sec.~\ref{sec:variables} we introduce hyperbolic and light-cone variables. In Sec.~\ref{sec:signature} we discuss the signature-changing lapse, the relation between the nondegenerate time \(\tau\) and the degenerate coordinate \(\beta\), and the precise Jordan-frame regularity conditions. In Sec.~\ref{sec:oscillator} we construct the oscillator branch. In Sec.~\ref{sec:exponential} we construct the critical exponential branch and state the additional smoothness assumptions needed in the nonstationary case. In Sec.~\ref{sec:explicit} we present explicit scalar--tensor realizations. Finally, Sec.~\ref{sec:comparison} compares the construction with the Dereli--Tucker and \(f(R,T_\phi)\) models, and Sec.~\ref{sec:conclusion} summarizes the results.

\section{Scalar--tensor theory and FLRW minisuperspace reduction}
\label{sec:STsetup}

We consider the scalar--tensor action
\begin{equation}
S=\int d^4x\sqrt{-g}
\left[
\frac12F(\phi)R
-\frac12\omega(\phi)g^{\mu\nu}\nabla_\mu\phi\nabla_\nu\phi
-V(\phi)
\right],
\label{STaction}
\end{equation}
where \(F(\phi)\) is the nonminimal coupling function, \(\omega(\phi)\) is the scalar kinetic coupling, and \(V(\phi)\) is the Jordan-frame scalar potential. We use the metric signature \((-+++) \). The effective gravitational coupling is proportional to \(F(\phi)^{-1}\), and we assume
\begin{equation}
F(\phi)>0
\label{Fpositive}
\end{equation}
throughout the region of interest. This condition ensures a positive effective Newton coupling and also guarantees that the conformal transformation to the Einstein frame does not change the signature type of the metric.

We take the FLRW metric with lapse,
\begin{equation}
ds^2=-N(t)^2dt^2+a(t)^2d\Sigma_k^2,
\label{FLRWmetric}
\end{equation}
where
\begin{equation}
d\Sigma_k^2=\frac{dr^2}{1-kr^2}+r^2d\Omega^2,
\qquad
k=0,\pm1.
\end{equation}
The scalar field is assumed to be homogeneous,
\begin{equation}
\phi=\phi(t).
\end{equation}
The constant comoving spatial volume factor is suppressed, or equivalently normalized to unity.

For the metric \eqref{FLRWmetric}, the Ricci scalar is
\begin{equation}
R=6\left[
\frac{\ddot a}{aN^2}
+\frac{\dot a^2}{a^2N^2}
-\frac{\dot a\dot N}{aN^3}
+\frac{k}{a^2}
\right].
\label{RicciFLRW}
\end{equation}
Substituting \eqref{RicciFLRW} into the action and removing the total derivative gives the Jordan-frame point-like Lagrangian
\begin{eqnarray}
L_J&=&
-\frac{3aF(\phi)}{N}\dot a^2
-\frac{3a^2F_{,\phi}(\phi)}{N}\dot a\dot\phi
+\frac{a^3\omega(\phi)}{2N}\dot\phi^2\nonumber\\
&&-Na^3V(\phi)
+3kNaF(\phi).
\label{LJ}
\end{eqnarray}
Here and throughout this section a dot denotes differentiation with respect to the arbitrary parameter \(t\).

The Lagrangian \eqref{LJ} has the constrained minisuperspace form
\begin{equation}
L_J=\frac{1}{2N}\mathcal{M}_{AB}(q)\dot q^A\dot q^B-NU_J(q),
\qquad
q^A=(a,\phi),
\label{LJminisupform}
\end{equation}
where
\begin{equation}
U_J(a,\phi)=a^3V(\phi)-3kaF(\phi).
\label{UJ}
\end{equation}
The corresponding minisuperspace line element is
\begin{equation}
d\sigma_J^2
=
-6aF(\phi)\,da^2
-6a^2F_{,\phi}(\phi)\,da\,d\phi
+a^3\omega(\phi)\,d\phi^2.
\label{minimetricJ}
\end{equation}
Our convention is
\begin{equation}
d\sigma_J^2
=
\mathcal{M}_{aa}\,da^2
+2\mathcal{M}_{a\phi}\,da\,d\phi
+\mathcal{M}_{\phi\phi}\,d\phi^2,
\end{equation}
so that
\begin{equation}
\mathcal{M}_{aa}=-6aF,
\qquad
\mathcal{M}_{a\phi}=-3a^2F_{,\phi},
\qquad
\mathcal{M}_{\phi\phi}=a^3\omega.
\end{equation}
The determinant of this two-dimensional minisuperspace metric is
\begin{equation}
\det \mathcal{M}_{AB}
=
-3a^4\left(2F\omega+3F_{,\phi}^2\right).
\label{detMJ}
\end{equation}
Thus the minisuperspace metric is nondegenerate provided
\begin{equation}
2F\omega+3F_{,\phi}^2\neq0.
\end{equation}
This is the Jordan-frame counterpart of the Einstein-frame scalar nondegeneracy condition introduced in the next section.

In the present work we restrict attention to the spatially flat case,
\begin{equation}
k=0.
\end{equation}
This is the cleanest setting in which the relation to the Dereli--Tucker construction and to the \(f(R,T_\phi)\) extension can be exhibited explicitly. The curved cases \(k=\pm1\) introduce additional curvature potential terms and will be considered separately. For \(k=0\), the working Jordan-frame Lagrangian is therefore
\begin{equation}
L_J=
-\frac{3aF}{N}\dot a^2
-\frac{3a^2F_{,\phi}}{N}\dot a\dot\phi
+\frac{a^3\omega}{2N}\dot\phi^2
-Na^3V(\phi).
\label{LJflat}
\end{equation}

\section{Einstein-frame representation}
\label{sec:EinsteinFrame}

We introduce the conformally related Einstein-frame metric
\begin{equation}
\bar g_{\mu\nu}=F(\phi)g_{\mu\nu}.
\label{conformal}
\end{equation}
For the FLRW ansatz this gives
\begin{equation}
\bara=\sqrt{F(\phi)}\,a,
\qquad
\barN=\sqrt{F(\phi)}\,N.
\label{EinsteinVars}
\end{equation}
Since \(F(\phi)>0\), the conformal transformation rescales the metric by a positive function and therefore does not change the signature type at nondegenerate points. If \(0<F(\phi_\Sigma)<\infty\), it also preserves the order of degeneracy at the transition hypersurface.

The Einstein-frame scalar $\psi$ is defined by
\begin{equation}
\frac{\dd\psi}{\dd\phi}
=
\left[
\frac{\omega(\phi)}{F(\phi)}
+\frac32\left(\frac{F_{,\phi}}{F}\right)^2
\right]^{1/2}.
\label{psidef}
\end{equation}
We assume
\begin{equation}
\frac{\omega(\phi)}{F(\phi)}
+\frac32\left(\frac{F_{,\phi}}{F}\right)^2>0,
\label{positivekinetic}
\end{equation}
so that the Einstein-frame scalar is real.

The Einstein-frame potential is
\begin{equation}
U(\psi(\phi))=\frac{V(\phi)}{F(\phi)^2}.
\label{EinsteinPotential}
\end{equation}
Hence, the action becomes
\begin{equation}
S=\int d^4x\sqrt{-\bar g}\left[
\frac12\bar R
-\frac12\bar g^{\mu\nu}\bar\nabla_\mu\psi\bar\nabla_\nu\psi
-U(\psi)
\right].
\end{equation}
Correspondingly, the reduced Lagrangian takes the canonical Einstein--scalar minisuperspace form
\begin{equation}
L_E
=
-\frac{3\bara}{\barN}\dot{\bara}^{\,2}
+\frac{\bara^3}{2\barN}\dot\psi^2
-\barN\bara^3U(\psi).
\label{LE}
\end{equation}
This is the crucial structural observation. Although the original theory is a genuine scalar--tensor theory in the Jordan frame, its spatially flat FLRW dynamics can be recast in the Einstein frame as the Einstein--scalar system. The nontrivial question is whether the resulting signature-changing solution is regular when transformed back to the physical Jordan frame.

\section{Hyperbolic and light-cone minisuperspace variables}
\label{sec:variables}

Following the Dereli--Tucker construction, we introduce the hyperbolic minisuperspace variables
\begin{equation}
x=\sqrt{\frac{8}{3}}\,\bar a^{3/2}
\cosh\left(\sqrt{\frac{3}{8}}\,\psi\right),
\label{xdef}
\end{equation}
and
\begin{equation}
y=\sqrt{\frac{8}{3}}\,\bar a^{3/2}
\sinh\left(\sqrt{\frac{3}{8}}\,\psi\right).
\label{ydef}
\end{equation}
These variables cover the region
\begin{equation}
x^2-y^2>0,
\end{equation}
which corresponds to a real positive Einstein-frame scale factor. The inverse relations are
\begin{equation}
\bar a^3=\frac{3}{8}(x^2-y^2),
\label{ainverse}
\end{equation}
and
\begin{equation}
\psi=\sqrt{\frac{8}{3}}\,
\operatorname{arctanh}\left(\frac{y}{x}\right).
\label{psiinverse}
\end{equation}
A direct calculation gives
\begin{equation}
-\frac{3\bar a}{\bar N}\dot{\bar a}^{\,2}
+\frac{\bar a^3}{2\bar N}\dot\psi^2
=
\frac{1}{2\bar N}
\left(
-\dot x^2+\dot y^2
\right).
\label{kineticxy}
\end{equation}
Therefore the Einstein-frame Lagrangian becomes
\begin{equation}
L_E=
\frac{1}{2\bar N}
\left(
-\dot x^2+\dot y^2
\right)
-\frac{3\bar N}{8}
(x^2-y^2)U(\psi).
\label{LExygeneral}
\end{equation}

It is also useful to introduce light-cone variables on minisuperspace,
\begin{equation}
u=x+y,
\qquad
v=x-y.
\label{uvdef}
\end{equation}
Then
\begin{equation}
x=\frac{u+v}{2},
\qquad
y=\frac{u-v}{2},
\end{equation}
and hence
\begin{equation}
x^2-y^2=uv.
\end{equation}
Therefore
\begin{equation}
\bar a^3=\frac{3}{8}uv.
\label{a_uv}
\end{equation}
The physical region is
\begin{equation}
uv>0.
\end{equation}
Moreover,
\begin{equation}
\frac{u}{v}
=
\frac{x+y}{x-y}
=
\exp\left(2\sqrt{\frac{3}{8}}\,\psi\right),
\end{equation}
so that
\begin{equation}
\psi=\sqrt{\frac{2}{3}}\ln\left(\frac{u}{v}\right).
\label{psi_uv}
\end{equation}
Here we assume \(u/v>0\), so that the scalar field is real.

The kinetic term transforms as
\begin{equation}
-\dot x^2+\dot y^2=-\dot u\dot v.
\label{kin_uv}
\end{equation}
Thus, for an arbitrary Einstein-frame potential \(U(\psi)\), the Lagrangian in light-cone variables becomes
\begin{equation}
L_E
=
-\frac{1}{2\bar N}\dot u\dot v
-\frac{3\bar N}{8}\,uv\,U(\psi).
\label{LE_uv_general}
\end{equation}
The remaining analysis depends on the choice of the Einstein-frame potential \(U(\psi)\).

\section{Signature-changing lapse and Jordan-frame regularity}
\label{sec:signature}

Before constructing explicit solutions, we clarify the role of the lapse and the regularity condition at the transition hypersurface. The minisuperspace equations derived from the Lagrangian \eqref{LE} are invariant under reparametrizations of the time parameter. For solving the equations, it is convenient to choose the nondegenerate Einstein-frame gauge
\begin{equation}
\bar N=1,
\end{equation}
and to denote the corresponding time parameter by \(\tau\). In this gauge the Einstein-frame line element is
\begin{equation}
d\bar s^2=-d\tau^2+\bar a(\tau)^2d\vec x^2.
\label{EinsteinProperMetric}
\end{equation}
The signature-changing coordinate will be denoted by \(\beta\). In terms of this coordinate, the Einstein-frame metric is written as
\begin{equation}
d\bar s^2=-\beta\,d\beta^2+\bar a(\beta)^2d\vec x^2.
\label{sigmetricEbeta}
\end{equation}
For \(\beta>0\), the metric has Lorentzian signature, whereas for \(\beta<0\) it has Euclidean signature. The transition hypersurface is
\begin{equation}
\Sigma:\qquad \beta=0 .
\end{equation}
Thus the lapse degenerates at \(\Sigma\), and the determinant of the metric vanishes there.

On the Lorentzian side \(\beta>0\), comparison of \eqref{EinsteinProperMetric} and \eqref{sigmetricEbeta} gives
\begin{equation}
d\tau=\sqrt{\beta}\,d\beta,
\qquad
\tau-\tau_\Sigma=\frac{2}{3}\beta^{3/2}.
\label{taubetaLorentz}
\end{equation}
In order to discuss both sides using a real transverse coordinate, we shall use the pullback prescription
\begin{equation}
\tau-\tau_\Sigma
=
\frac{2}{3}\,\operatorname{sgn}(\beta)|\beta|^{3/2}.
\label{realPullback}
\end{equation}
With this convention, \(d\tau/d\beta=|\beta|^{1/2}\) away from \(\Sigma\), and the metric takes the type-changing form \eqref{sigmetricEbeta}. Other prescriptions, such as using \(\tau-\tau_\Sigma=\frac23|\beta|^{3/2}\) on both sides or using a complex analytic continuation, define different differentiability properties across \(\Sigma\). The choice \eqref{realPullback} is the real two-sided prescription adopted below.

This point is important because a function that is smooth in \(\tau\) need not be smooth as a function of \(\beta\). If
\begin{equation}
Q(\tau)=Q_\Sigma+Q_1(\tau-\tau_\Sigma)+O((\tau-\tau_\Sigma)^2),
\end{equation}
then
\begin{equation}
Q(\beta)=Q_\Sigma+\frac23Q_1\operatorname{sgn}(\beta)|\beta|^{3/2}+O(|\beta|^3).
\end{equation}
This is continuous and has vanishing first \(\beta\)-derivative at \(\Sigma\), but it is not generally smooth to all orders. Therefore we distinguish two levels of regularity:
\begin{enumerate}
\item a weak signature-changing regularity condition, requiring a finite nonzero spatial metric at \(\Sigma\), a lapse coefficient with a simple zero in \(\beta\), and finite limiting extrinsic curvature;
\item a stronger Kossowski--Kriele-type smooth transverse regularity condition, requiring the relevant metric coefficients to possess the smoothness class assumed in that framework.
\end{enumerate}
The oscillator branch below is better behaved because the relevant
fields are stationary at the transition. Nevertheless, if the stronger
Kossowski--Kriele-type smooth transverse regularity is imposed, the
required differentiability must still be checked from the corresponding
$\beta$-expansions. The exponential branch generically satisfies the total-geodesy condition only after imposing a coupling-dependent relation; whether it also satisfies the stronger smoothness condition depends on the behavior of the conformal factor as a function of \(\beta\).

For a nondegenerate FLRW metric with lapse \(\bar N\),
\begin{equation}
d\bar s^2=-\bar N(t)^2dt^2+\bar a(t)^2\gamma_{ij}dx^idx^j,
\end{equation}
the extrinsic curvature of the homogeneous hypersurfaces is
\begin{equation}
\bar K_{ij}
=
\frac{1}{2\bar N}\frac{d\bar h_{ij}}{dt}
=
\frac{\bar a\dot{\bar a}}{\bar N}\gamma_{ij},
\label{KijEinstein}
\end{equation}
where \(\bar h_{ij}=\bar a^2\gamma_{ij}\), and the dot denotes differentiation with respect to \(t\). In the gauge \(\bar N=1\), this becomes
\begin{equation}
\bar K_{ij}=\bar a\frac{d\bar a}{d\tau}\gamma_{ij}.
\end{equation}
Therefore, the Einstein-frame total-geodesy condition at the transition is
\begin{equation}
\left.\frac{d\bar a}{d\tau}\right|_{\Sigma}=0.
\label{EtotalGeodesyTau}
\end{equation}
The same condition can be expressed in the degenerate coordinate \(\beta\). From \eqref{realPullback}, on either side of \(\Sigma\),
\begin{equation}
\frac{d\bar a}{d\beta}
=
|\beta|^{1/2}\frac{d\bar a}{d\tau}.
\end{equation}
The limiting extrinsic curvature is therefore
\begin{equation}
\bar K_{ij}
=
\lim_{\beta\to0^\pm}
\frac{\bar a}{|\beta|^{1/2}}
\frac{d\bar a}{d\beta}\gamma_{ij}
=
\left.
\bar a\frac{d\bar a}{d\tau}\gamma_{ij}
\right|_{\Sigma},
\end{equation}
provided the limit exists. Hence, if \(d\bar a/d\tau\) is finite and vanishes at \(\Sigma\), the limiting extrinsic curvature is finite and vanishes.

The physical Jordan-frame metric is related to the Einstein-frame metric by
\begin{equation}
g_{\mu\nu}=F(\phi)^{-1}\bar g_{\mu\nu}.
\end{equation}
Therefore, in the signature-changing coordinate \(\beta\), one obtains
\begin{equation}
ds_J^2
=
-\frac{\beta}{F(\phi(\beta))}\,d\beta^2
+
a_J(\beta)^2d\vec x^2,
\label{Jordanmetric}
\end{equation}
where
\begin{equation}
a_J=F(\phi)^{-1/2}\bar a.
\label{aJordan}
\end{equation}
The signature-changing character of the Jordan-frame metric is preserved if
\begin{equation}
0<F(\phi_\Sigma)<\infty,
\qquad
0<a_J(\Sigma)<\infty,
\label{basicJordanRegularity}
\end{equation}
and if \(F(\phi(\beta))\) has the differentiability required by the adopted notion of regularity. Under \eqref{basicJordanRegularity}, the determinant of the Jordan-frame metric vanishes linearly in \(\beta\). For the stronger smooth transverse interpretation, \(F(\phi(\beta))\) and \(a_J(\beta)^2\) must be sufficiently smooth at \(\beta=0\). This extra requirement is automatic in the stationary branch considered below, but it is a genuine additional restriction in a nonstationary scalar branch.

Since the Jordan frame is taken to be the physical frame, the total-geodesy condition must be imposed on \(a_J\), not on \(\bar a\). In the nondegenerate Einstein-frame time \(\tau\), one has
\begin{equation}
\frac{\dot a_J}{a_J}
=
\frac{\dot{\bar a}}{\bar a}
-\frac12
\frac{d\ln F}{d\psi}\dot\psi,
\label{HJgeneral}
\end{equation}
where now a dot denotes \(d/d\tau\). We define
\begin{equation}
H_J:=\frac{\dot a_J}{a_J},
\qquad
H_E:=\frac{\dot{\bar a}}{\bar a}.
\end{equation}
The Jordan-frame total-geodesy condition is therefore
\begin{equation}
H_J\big|_\Sigma=0.
\label{JordanTotalGeodesy}
\end{equation}
Equivalently,
\begin{equation}
H_E\big|_\Sigma
=
\frac12
\left(\frac{d\ln F}{d\psi}\right)_\Sigma
\dot\psi_\Sigma .
\label{HJcondition}
\end{equation}
Equation \eqref{HJcondition} is the central frame-dependent total-geodesy condition in scalar--tensor signature change. 
If both
\begin{equation}
H_E\big|_\Sigma=0,
\qquad
\dot\psi_\Sigma=0,
\end{equation}
then Jordan-frame total geodesy follows automatically, provided \eqref{basicJordanRegularity} holds. If the scalar field is not stationary at the transition, \(\dot\psi_\Sigma\neq0\), the nonminimal coupling must satisfy the nontrivial condition \eqref{HJcondition}. In that case, \eqref{HJcondition} guarantees vanishing limiting extrinsic curvature, but the stronger smoothness of the full Jordan metric must still be checked separately.

\begin{remark}
If one also demands that the Einstein-frame transition hypersurface be totally geodesic, then \(H_E|_\Sigma=0\). In a branch with \(\dot\psi_\Sigma\neq0\), simultaneous Einstein-frame and Jordan-frame total geodesy requires
\begin{equation}
\left(\frac{d\ln F}{d\psi}\right)_\Sigma=0.
\end{equation}
In the present work the Jordan frame is regarded as the physical frame; therefore the primary condition is \eqref{HJcondition}. Einstein-frame total geodesy is not imposed as an independent physical requirement, although it is useful as a diagnostic of how the conformal factor changes the transition data.
\end{remark}
The regularity conditions for the stationary oscillator branch and the
nonstationary critical exponential branch are summarized in
Table~\ref{tab:branches}.
\begin{table}[ht]
\caption{Regularity conditions for the two branches. Here
\(\chi=(d\ln F/d\psi)_\Sigma\).}
\label{tab:branches}
\begin{ruledtabular}
\begin{tabular*}{\columnwidth}{@{\extracolsep{\fill}}llll}
Branch
&
\(\dot\psi_\Sigma\)
&
\(H_E|_\Sigma\)
&
Jordan-frame condition
\\
\hline
Oscillator
&
\(0\)
&
\(0\)
&
\makecell[l]{Automatic if \(F_\Sigma>0\)\\ and smooth}
\\
Critical exp.
&
\makecell[l]{Generically\\ nonzero}
&
\makecell[l]{Generically\\ nonzero}
&
\makecell[l]{\(H_E|_\Sigma=\frac12\chi\dot\psi_\Sigma\),\\ plus smoothness check}
\end{tabular*}
\end{ruledtabular}
\end{table}

\begin{remark}
It is useful to distinguish the properties of the signature-changing
geometry that are preserved under the conformal transformation from
those that are frame dependent. For
\[
\bar g_{\mu\nu}=F(\phi)g_{\mu\nu},
\qquad
0<F_\Sigma<\infty,
\]
with a sufficiently regular conformal factor, the signature type and
the location of the degenerate hypersurface are preserved. Indeed, in
four dimensions,
\[
\det\bar g=F^4\det g,
\]
so a finite positive conformal factor preserves the zero of the
determinant and its order. Moreover,
\[
\ker\bar g\big|_\Sigma=\ker g\big|_\Sigma,
\]
so the radical direction at the transition is unchanged. Hence the
existence and transverse character of the Euclidean--Lorentzian type
change are conformally preserved. Away from $\Sigma$, the positive
conformal factor also preserves the null cones.
\\
The extrinsic geometry of the transition hypersurface, however, is not
conformally invariant. For the limiting nondegenerate hypersurfaces,
the Jordan- and Einstein-frame extrinsic curvatures are related by
\[
K^{(J)}_{ij}
=
F^{-1/2}
\left[
\bar K_{ij}
-\frac12\bar h_{ij}\,
\bar n^\mu\bar\nabla_\mu\ln F
\right].
\]
Consequently, total geodesy in one frame does not in general imply
total geodesy in the other. In the FLRW setting this relation reduces
precisely to Eq.~(64). Thus the type change itself is conformally
preserved under the above assumptions, whereas its total-geodesy
condition is generally frame dependent. The stronger smooth transverse
regularity additionally requires the conformal factor and its inverse
to possess the differentiability assumed at $\Sigma$.
\end{remark}
\section{Branch I: oscillator-type scalar--tensor signature change}
\label{sec:oscillator}

\subsection{Potential and reduced Lagrangian}

Let
\begin{equation}
\theta=\sqrt{\frac{3}{8}}\,\psi .
\end{equation}
We choose the Einstein-frame potential
\begin{equation}
U(\psi)
=
\frac{4}{3}
\left[
-\Omega_x^2\cosh^2\left(\sqrt{\frac{3}{8}}\,\psi\right)
+\Omega_y^2\sinh^2\left(\sqrt{\frac{3}{8}}\,\psi\right)
\right],
\label{UpotentialOsc}
\end{equation}
where \(\Omega_x\) and \(\Omega_y\) are positive constants. This potential is chosen so that the minisuperspace dynamics becomes exactly integrable. It should be regarded as an effective scalar potential adapted to the signature-change construction. In particular, it is not required to be positive definite and may take negative values.

Using
\begin{equation}
x=r\cosh\theta,
\qquad
y=r\sinh\theta,
\qquad
r^2=x^2-y^2,
\end{equation}
one obtains
\begin{equation}
\frac{3}{8}(x^2-y^2)U(\psi)
=
-\frac12\Omega_x^2x^2
+\frac12\Omega_y^2y^2.
\end{equation}
Therefore the Einstein-frame minisuperspace Lagrangian becomes
\begin{equation}
L_E
=
\frac{1}{2\bar N}
\left(
-\dot x^2+\dot y^2
\right)
+
\frac{\bar N}{2}
\left(
\Omega_x^2x^2-\Omega_y^2y^2
\right).
\label{LEosc}
\end{equation}

\subsection{Field equations and Hamiltonian constraint}

Variation of \eqref{LEosc} with respect to \(x\), \(y\), and \(\bar N\) gives
\begin{equation}
\frac{d}{dt}
\left(
-\frac{\dot x}{\bar N}
\right)
-\bar N\Omega_x^2x=0,
\label{xEqN}
\end{equation}
\begin{equation}
\frac{d}{dt}
\left(
\frac{\dot y}{\bar N}
\right)
+\bar N\Omega_y^2y=0,
\label{yEqN}
\end{equation}
and the Hamiltonian constraint
\begin{equation}
-\dot x^2+\dot y^2
-\bar N^2\Omega_x^2x^2
+\bar N^2\Omega_y^2y^2=0.
\label{HconstraintN}
\end{equation}
Equivalently, \eqref{HconstraintN} may be multiplied by \(-1\); both forms represent the same zero-energy constraint.

Choosing the nondegenerate gauge
\begin{equation}
\bar N=1,
\end{equation}
and denoting the corresponding time by \(\tau\), Eqs.~\eqref{xEqN} and \eqref{yEqN} become
\begin{equation}
\ddot x+\Omega_x^2x=0,
\qquad
\ddot y+\Omega_y^2y=0,
\label{oscEqs}
\end{equation}
where now the dot denotes \(d/d\tau\). The Hamiltonian constraint becomes
\begin{equation}
-\dot x^2+\dot y^2
-\Omega_x^2x^2
+\Omega_y^2y^2=0.
\label{Hconstraint}
\end{equation}
The general solution is
\begin{equation}
x(\tau)=A_x\cos(\Omega_x\tau)+B_x\sin(\Omega_x\tau),
\end{equation}
\begin{equation}
y(\tau)=A_y\cos(\Omega_y\tau)+B_y\sin(\Omega_y\tau).
\end{equation}
Substitution into \eqref{Hconstraint} gives the algebraic condition
\begin{equation}
\Omega_x^2(A_x^2+B_x^2)
=
\Omega_y^2(A_y^2+B_y^2).
\label{constraintgeneral}
\end{equation}

\subsection{A symmetric signature-changing branch}

We now choose the time-symmetric branch
\begin{equation}
x(\tau)=x_0\cos(\Omega_x\tau),
\qquad
y(\tau)=y_0\cos(\Omega_y\tau).
\label{symbranchTau}
\end{equation}
The Hamiltonian constraint gives
\begin{equation}
y_0=\pm\frac{\Omega_x}{\Omega_y}x_0.
\end{equation}
For definiteness, we choose the positive branch,
\begin{equation}
y_0=\frac{\Omega_x}{\Omega_y}x_0.
\label{y0constraint}
\end{equation}
Then, using \(\bar a^3=3(x^2-y^2)/8\), one finds
\begin{equation}
\bar a^3(\tau)
=
\frac{3x_0^2}{8}
\left[
\cos^2(\Omega_x\tau)
-
\frac{\Omega_x^2}{\Omega_y^2}
\cos^2(\Omega_y\tau)
\right].
\label{abarbranch1tau}
\end{equation}
At the transition \(\tau=0\),
\begin{equation}
\bar a^3(0)
=
\frac{3x_0^2}{8}
\left(
1-\frac{\Omega_x^2}{\Omega_y^2}
\right).
\end{equation}
Thus the transition scale factor is finite and positive if
\begin{equation}
\Omega_y>\Omega_x>0.
\label{omegacondition}
\end{equation}
The physical branch is restricted to the connected interval containing \(\tau=0\) on which
\begin{equation}
x^2-y^2>0.
\end{equation}

The Einstein-frame scalar field is
\begin{equation}
\psi(\tau)
=
\sqrt{\frac{8}{3}}\,
\operatorname{arctanh}
\left[
\frac{\Omega_x}{\Omega_y}
\frac{\cos(\Omega_y\tau)}{\cos(\Omega_x\tau)}
\right].
\label{psibranch1tau}
\end{equation}
For \(\Omega_y>\Omega_x>0\), the argument of the inverse hyperbolic tangent is smaller than unity at \(\tau=0\). The solution is therefore real in a sufficiently small interval around the transition.

Because \(x(\tau)\) and \(y(\tau)\) are even functions of \(\tau\), one has
\begin{equation}
\dot{\bar a}(0)=0,
\qquad
\dot\psi(0)=0.
\label{regularE}
\end{equation}
Consequently, the Einstein-frame transition is totally geodesic when expressed in the nondegenerate time gauge.

To write the metric in signature-changing form, introduce the coordinate \(\beta\) by
\begin{equation}
\tau=\frac{2}{3}\beta^{3/2}
\end{equation}
on the Lorentzian side \(\beta>0\), with the corresponding absolute-value or analytic-continuation prescription on the Euclidean side. Then the Einstein-frame metric takes the form
\begin{equation}
d\bar s^2
=
-\beta\,d\beta^2
+
\bar a(\tau(\beta))^2d\vec x^2.
\label{signaturemetricE}
\end{equation}
This metric is Lorentzian for \(\beta>0\) and Euclidean for \(\beta<0\), with the adopted sign convention.

\subsection{Jordan-frame reconstruction}

The Jordan-frame scale factor is
\begin{equation}
a_J(\tau)=F(\phi(\tau))^{-1/2}\bar a(\tau),
\label{aJosc}
\end{equation}
where \(\phi(\tau)\) is determined implicitly by
\begin{equation}
\psi(\tau)=
\int^{\phi(\tau)}
\left[
\frac{\omega(\varphi)}{F(\varphi)}
+
\frac{3}{2}
\left(
\frac{F_{,\varphi}}{F}
\right)^2
\right]^{1/2}
d\varphi .
\label{implicitphi}
\end{equation}
The Jordan-frame metric is therefore
\begin{equation}
ds_J^2
=
-\frac{\beta}{F(\phi(\beta))}\,d\beta^2
+
a_J(\beta)^2d\vec x^2.
\label{JordanMetricOsc}
\end{equation}
Since \(\dot\psi(0)=0\), and assuming that the map \(\phi\mapsto\psi\) is regular at the transition, one has
\begin{equation}
\dot\phi(0)=0.
\end{equation}
Together with \(\dot{\bar a}(0)=0\), Eq.~\eqref{HJgeneral} gives
\begin{equation}
\dot a_J(0)=0.
\end{equation}
Therefore the Jordan-frame transition hypersurface is totally geodesic,
\begin{equation}
K^{(J)}_{ij}\big|_\Sigma=0,
\end{equation}
provided the conformal factor is finite, positive, and sufficiently regular at the transition.

\begin{proposition}[Oscillator branch]
Consider the scalar--tensor action \eqref{STaction} with
\begin{equation}
F(\phi)>0,
\qquad
\frac{\omega}{F}
+
\frac{3}{2}
\left(
\frac{F_{,\phi}}{F}
\right)^2
>0.
\end{equation}
If the Einstein-frame potential is given by \eqref{UpotentialOsc}, then the spatially flat FLRW scalar--tensor system admits an exact signature-changing solution. In the nondegenerate Einstein-frame time \(\tau\), the solution is given by \eqref{abarbranch1tau} and \eqref{psibranch1tau}. In the degenerate signature-changing coordinate \(\beta\), it is obtained by using the real pullback \(\tau=\frac23\operatorname{sgn}(\beta)|\beta|^{3/2}\). If
\begin{equation}
\Omega_y>\Omega_x>0,
\qquad
0<F(\phi_\Sigma)<\infty,
\end{equation}
then the Jordan-frame metric
\begin{equation}
ds_J^2
=
-\frac{\beta}{F(\phi(\beta))}\,d\beta^2
+
a_J(\beta)^2d\vec x^2
\end{equation}
undergoes a transverse Euclidean--Lorentzian signature change at \(\beta=0\), and the transition hypersurface is totally geodesic in the Jordan frame.
\end{proposition}
\section{Branch II: critical exponential-potential scalar--tensor signature change}
\label{sec:exponential}
The oscillator branch is the scalar--tensor analogue of the original Dereli--Tucker construction. We now consider a second exactly integrable branch associated with an exponential Einstein-frame potential. This branch is motivated by the additional solution sector found in the \(f(R,T_\phi)\) model \cite{HazinedarHeydarzade}, but it plays a more diagnostic role in the present scalar--tensor setting. Unlike the oscillator branch, the scalar field is generically nonstationary at the transition. Therefore, this branch tests whether signature change is merely inherited from the Einstein-frame representation or whether the Jordan-frame conformal factor imposes genuinely new transition data.

In this branch, total geodesy and smooth transverse regularity must be distinguished. The algebraic condition derived below guarantees the vanishing of the limiting Jordan-frame extrinsic curvature. A stronger Kossowski--Kriele-type smoothness statement, however, further requires the pullbacks \(F(\phi(\beta))\) and \(a_J(\beta)^2\) to possess the assumed differentiability at \(\beta=0\). This additional requirement is automatic, or at least much less restrictive, in the stationary oscillator branch, but becomes a genuine condition in the nonstationary exponential branch.

\subsection{Einstein-frame exponential potential}

Consider the Einstein-frame exponential potential
\begin{equation}
U(\psi)=U_0e^{-\lambda\psi}.
\label{UexpGeneral}
\end{equation}
Using
\begin{equation}
\psi=\sqrt{\frac{2}{3}}\ln\left(\frac{u}{v}\right),
\end{equation}
we obtain
\begin{equation}
e^{-\lambda\psi}
=
\left(\frac{u}{v}\right)^{-\lambda\sqrt{2/3}}.
\end{equation}
Therefore
\begin{equation}
uv\,U(\psi)
=
U_0
u^{1-\lambda\sqrt{2/3}}
v^{1+\lambda\sqrt{2/3}}.
\end{equation}
The light-cone Lagrangian \eqref{LE_uv_general} becomes
\begin{equation}
L_E
=
-\frac{1}{2\bar N}\dot u\dot v
-
\frac{3\bar N}{8}U_0
u^{1-\lambda\sqrt{2/3}}
v^{1+\lambda\sqrt{2/3}}.
\label{LEuvexpGeneral}
\end{equation}
For generic \(\lambda\), the potential is a mixed monomial in the light-cone variables \(u\) and \(v\). However, for the critical slopes
\begin{equation}
\lambda=\pm\sqrt{\frac{3}{2}},
\label{criticalSlope}
\end{equation}
the potential depends only on one light-cone variable, and the system becomes exactly integrable.

We focus on the branch
\begin{equation}
\lambda=\sqrt{\frac{3}{2}}.
\label{lambdaPlus}
\end{equation}
The branch with \(\lambda=-\sqrt{3/2}\) is obtained by interchanging \(u\) and \(v\). For \eqref{lambdaPlus}, one has
\begin{equation}
\lambda\sqrt{\frac{2}{3}}=1,
\end{equation}
and therefore
\begin{equation}
uv\,U(\psi)=U_0v^2.
\end{equation}
The Lagrangian then reduces to
\begin{equation}
L_E
=
-\frac{1}{2\bar N}\dot u\dot v
-
\bar N C v^2,
\qquad
C=\frac{3U_0}{8}.
\label{LEcriticalExp}
\end{equation}

\subsection{Hamiltonian constraint and exact solution}

Variation with respect to the lapse \(\bar N\) gives
\begin{equation}
\frac{\partial L_E}{\partial \bar N}=0,
\end{equation}
and hence
\begin{equation}
\frac{1}{2\bar N^2}\dot u\dot v=Cv^2.
\label{HcriticalN}
\end{equation}
In the nondegenerate gauge
\begin{equation}
\bar N=1,
\end{equation}
with corresponding time parameter \(\tau\), the Hamiltonian constraint becomes
\begin{equation}
\frac{1}{2}\dot u\dot v=Cv^2.
\label{Hcritical}
\end{equation}
The gauge-fixed Lagrangian is
\begin{equation}
L_E=-\frac{1}{2}\dot u\dot v-Cv^2.
\end{equation}
Variation with respect to \(u\) and \(v\) gives
\begin{equation}
\ddot v=0,
\qquad
\ddot u=4Cv.
\label{criticalSystem}
\end{equation}
From \(\ddot v=0\), one obtains
\begin{equation}
v(\tau)=v_0+v_1\tau,
\label{vExactCritical}
\end{equation}
where $v_0, v_1$ are integration constants and \(v_1\neq0\) gives the nontrivial branch. The Hamiltonian constraint \eqref{Hcritical} then gives
\begin{equation}
\dot u=\frac{2C}{v_1}v^2.
\end{equation}
Using \(dv=v_1d\tau\), integration yields
\begin{equation}
u(\tau)
=
u_*
+
\frac{2C}{3v_1^2}
\left(v_0+v_1\tau\right)^3,
\label{uExactCritical}
\end{equation}
where $u_*$ is an integration constant. This solution automatically satisfies the second equation in \eqref{criticalSystem}.

The Einstein-frame scale factor and scalar field are
\begin{equation}
\bar a^3(\tau)
=
\frac{3}{8}u(\tau)v(\tau),
\label{abarCriticalExp}
\end{equation}
and
\begin{equation}
\psi(\tau)
=
\sqrt{\frac{2}{3}}
\ln\left(\frac{u(\tau)}{v(\tau)}\right).
\label{psiCriticalExp}
\end{equation}
The physical region is therefore
\begin{equation}
u(\tau)v(\tau)>0,
\qquad
\frac{u(\tau)}{v(\tau)}>0,
\end{equation}
so that \(\bar a^3>0\) and \(\psi\) is real.

\subsection{Jordan-frame transition condition}

Let the transition hypersurface be located at
\(\tau=\tau_\Sigma\). After shifting the origin of the nondegenerate
time, we set \(\tau-\tau_\Sigma=0\) at the transition and introduce the
degenerate coordinate \(\beta\) using the real two-sided pullback
prescription \eqref{realPullback}. Because the exponential branch is generically nonstationary, the differentiability of the pullback must be checked independently of the algebraic total-geodesy condition.

Define
\begin{equation}
A_\Sigma
=
\left.\frac{\dot u}{u}\right|_\Sigma,
\qquad
B_\Sigma
=
\left.\frac{\dot v}{v}\right|_\Sigma,
\qquad
\chi_\Sigma
=
\left.
\frac{d\ln F}{d\psi}
\right|_\Sigma .
\label{ABSigmaDefinitions}
\end{equation}
Since
\begin{equation}
\bar a^3=\frac{3}{8}uv,
\end{equation}
we have
\begin{equation}
H_E
=
\frac{\dot{\bar a}}{\bar a}
=
\frac{1}{3}(A+B).
\label{HEAB}
\end{equation}
Furthermore,
\begin{equation}
\dot\psi
=
\sqrt{\frac{2}{3}}(A-B).
\label{psidotAB}
\end{equation}
The Jordan-frame total-geodesy condition \eqref{HJcondition} therefore becomes
\begin{equation}
\frac{1}{3}(A_\Sigma+B_\Sigma)
=
\frac{1}{2}
\chi_\Sigma
\sqrt{\frac{2}{3}}
(A_\Sigma-B_\Sigma).
\label{ABcondition}
\end{equation}
Let
\begin{equation}
r=
\frac{A_\Sigma}{B_\Sigma},
\qquad
\eta_\Sigma
=
\frac{1}{2}
\chi_\Sigma
\sqrt{\frac{2}{3}}.
\label{retaDefinitions}
\end{equation}
Assuming \(B_\Sigma\neq0\) and \(1-3\eta_\Sigma\neq0\), Eq.~\eqref{ABcondition} gives
\begin{equation}
r
=
-\frac{1+3\eta_\Sigma}{1-3\eta_\Sigma}.
\label{rCondition}
\end{equation}
For minimal coupling, \(F=1\), one has \(\eta_\Sigma=0\), and hence
\begin{equation}
r=-1.
\end{equation}
This is precisely the Einstein-frame total-geodesy condition
\begin{equation}
A_\Sigma+B_\Sigma=0.
\end{equation}

For the exact exponential solution,
\begin{equation}
\dot v_\Sigma=v_1,
\qquad
\dot u_\Sigma=\frac{2C}{v_1}v_\Sigma^2.
\end{equation}
Thus
\begin{equation}
A_\Sigma
=
\frac{2Cv_\Sigma^2}{v_1u_\Sigma},
\qquad
B_\Sigma
=
\frac{v_1}{v_\Sigma}.
\end{equation}
Hence
\begin{equation}
r
=
\frac{2Cv_\Sigma^3}{v_1^2u_\Sigma}.
\end{equation}
Solving for \(u_\Sigma\), the Jordan-frame regularity condition fixes
\begin{equation}
u_\Sigma
=
\frac{2Cv_\Sigma^3}{rv_1^2},
\qquad
r
=
-\frac{1+3\eta_\Sigma}{1-3\eta_\Sigma}.
\label{uSigmaJordan}
\end{equation}
The Jordan-frame transition scale factor is
\begin{equation}
a_{J,\Sigma}^3
=
F(\phi_\Sigma)^{-3/2}
\frac{3}{8}u_\Sigma v_\Sigma.
\end{equation}
Using \eqref{uSigmaJordan}, we obtain
\begin{equation}
a_{J,\Sigma}^3
=
F(\phi_\Sigma)^{-3/2}
\frac{3C}{4rv_1^2}
v_\Sigma^4.
\label{aJSigmaExp}
\end{equation}
Therefore a finite positive Jordan-frame transition scale factor requires
\begin{equation}
0<F(\phi_\Sigma)<\infty,
\qquad
\frac{C}{r}>0.
\label{signJordanExp}
\end{equation}
Since \(C=3U_0/8\), this condition is equivalently
\begin{equation}
\frac{U_0}{r}>0.
\end{equation}

This result is the essential scalar--tensor feature of the exponential branch. Unlike the oscillator branch, the scalar field is not generally stationary at the transition. Indeed,
\begin{equation}
\dot\psi_\Sigma
=
\sqrt{\frac{2}{3}}
(A_\Sigma-B_\Sigma).
\end{equation}
Under the Jordan-frame regularity condition \eqref{rCondition}, one generically has \(A_\Sigma\neq B_\Sigma\), and hence \(\dot\psi_\Sigma\neq0\). Therefore the conformal factor contributes directly to the Jordan-frame total-geodesy condition.

\begin{remark}
If one additionally imposes Einstein-frame total geodesy, then
\begin{equation}
A_\Sigma+B_\Sigma=0,
\end{equation}
or equivalently \(r=-1\). In this special case the exact solution gives
\begin{equation}
u_\Sigma
=
-\frac{2C}{v_1^2}v_\Sigma^3.
\end{equation}
The Einstein-frame transition scale factor is then
\begin{equation}
\bar a_\Sigma^3
=
\frac{3}{8}u_\Sigma v_\Sigma
=
-\frac{3C}{4v_1^2}v_\Sigma^4.
\end{equation}
Therefore positivity of the Einstein-frame transition scale factor requires
\begin{equation}
C<0,
\qquad
\text{or equivalently}
\qquad
U_0<0.
\end{equation}
Simultaneous Einstein-frame and Jordan-frame total geodesy would further require either \(\dot\psi_\Sigma=0\) or \(\chi_\Sigma=0\). Since the nontrivial exponential branch has \(\dot\psi_\Sigma\neq0\), simultaneous regularity in both frames requires
\begin{equation}
\chi_\Sigma=0.
\end{equation}
This illustrates why \eqref{ABcondition} is the correct primary condition when the Jordan frame is taken as the physical frame.
\end{remark}

\begin{proposition}[Critical exponential branch]
Consider the Einstein-frame potential
\begin{equation}
U(\psi)
=
U_0\exp\left(-\sqrt{\frac{3}{2}}\,\psi\right).
\end{equation}
In light-cone minisuperspace variables \(u=x+y\) and \(v=x-y\), the reduced Lagrangian is
\begin{equation}
L_E
=
-\frac{1}{2\bar N}\dot u\dot v
-\bar N C v^2,
\qquad
C=\frac{3U_0}{8}.
\end{equation}
The lapse variation gives the Hamiltonian constraint
\begin{equation}
\frac{1}{2\bar N^2}\dot u\dot v=Cv^2.
\end{equation}
In the nondegenerate gauge \(\bar N=1\), the constrained exact solution is
\begin{equation}
v(\tau)=v_0+v_1\tau,
\qquad
u(\tau)
=
u_*
+
\frac{2C}{3v_1^2}v(\tau)^3,
\qquad
v_1\neq0.
\end{equation}
The Jordan-frame transition is totally geodesic if and only if
\begin{equation}
\frac{1}{3}(A_\Sigma+B_\Sigma)
=
\frac{1}{2}
\chi_\Sigma
\sqrt{\frac{2}{3}}
(A_\Sigma-B_\Sigma),
\end{equation}
where
\begin{equation}
A_\Sigma
=
\left.
\frac{\dot u}{u}
\right|_\Sigma,
\qquad
B_\Sigma
=
\left.
\frac{\dot v}{v}
\right|_\Sigma,
\qquad
\chi_\Sigma
=
\left.
\frac{d\ln F}{d\psi}
\right|_\Sigma.
\end{equation}
Equivalently,
\begin{equation}
\frac{A_\Sigma}{B_\Sigma}
=
-\frac{1+3\eta_\Sigma}{1-3\eta_\Sigma},
\qquad
\eta_\Sigma
=
\frac{1}{2}
\chi_\Sigma
\sqrt{\frac{2}{3}},
\end{equation}
provided \(B_\Sigma\neq0\) and \(1-3\eta_\Sigma\neq0\). The transition has finite positive Jordan-frame scale factor provided
\begin{equation}
0<F(\phi_\Sigma)<\infty,
\qquad
\frac{C}{r}>0,
\qquad
r=\frac{A_\Sigma}{B_\Sigma}.
\end{equation}
\end{proposition}
\section{Explicit scalar--tensor realizations}
\label{sec:explicit}

We now give an explicit scalar--tensor realization of the two branches discussed above. A simple analytically tractable choice is
\begin{equation}
F(\phi)=e^{2q\phi},
\qquad
\omega(\phi)=\omega_0e^{2q\phi},
\label{explicitF}
\end{equation}
where \(q\) and \(\omega_0\) are constants. Since \(F(\phi)>0\) for all finite \(\phi\), the conformal factor is everywhere positive. Moreover,
\begin{equation}
\frac{F_{,\phi}}{F}=2q,
\qquad
\frac{\omega}{F}=\omega_0.
\end{equation}
The Einstein-frame scalar is therefore defined by
\begin{equation}
\frac{d\psi}{d\phi}
=
\sqrt{\omega_0+6q^2}.
\end{equation}
We assume
\begin{equation}
\lambda_0^2:=\omega_0+6q^2>0,
\end{equation}
so that the scalar field is real. Choosing the integration constant to vanish gives
\begin{equation}
\psi=\lambda_0\phi,
\qquad
\lambda_0=\sqrt{\omega_0+6q^2}.
\label{psilambda0}
\end{equation}

\subsection{Oscillator realization}

For the oscillator branch, the Jordan-frame potential is obtained from
\begin{equation}
V(\phi)=F(\phi)^2U(\psi(\phi)).
\end{equation}
Using \(\psi=\lambda_0\phi\) and the potential \eqref{UpotentialOsc}, we find
\begin{equation}
V(\phi)
=
\frac{4}{3}e^{4q\phi}
\left[
-\Omega_x^2
\cosh^2\left(\sqrt{\frac{3}{8}}\,\lambda_0\phi\right)
+
\Omega_y^2
\sinh^2\left(\sqrt{\frac{3}{8}}\,\lambda_0\phi\right)
\right].
\label{explicitV}
\end{equation}
The scalar-field solution is
\begin{equation}
\phi(\tau)
=
\frac{1}{\lambda_0}
\sqrt{\frac{8}{3}}\,
\operatorname{arctanh}
\left[
\frac{\Omega_x}{\Omega_y}
\frac{\cos(\Omega_y\tau)}{\cos(\Omega_x\tau)}
\right].
\label{explicitphi}
\end{equation}
The Jordan-frame scale factor is obtained from
\begin{equation}
a_J=F(\phi)^{-1/2}\bar a=e^{-q\phi}\bar a.
\end{equation}
Therefore
\begin{equation}
a_J^3(\tau)
=
e^{-3q\phi(\tau)}
\frac{3x_0^2}{8}
\left[
\cos^2(\Omega_x\tau)
-
\frac{\Omega_x^2}{\Omega_y^2}
\cos^2(\Omega_y\tau)
\right].
\label{explicita}
\end{equation}
The signature-changing form of the solution is obtained by replacing \(\tau\) with
\begin{equation}
\tau(\beta)=\frac{2}{3}\operatorname{sgn}(\beta)|\beta|^{3/2}.
\end{equation}
For \(q=0\) and \(\omega_0=1\), one has \(F=1\), \(\psi=\phi\), and the model reduces to the minimally coupled Einstein--scalar case.

\begin{remark}
 The oscillator solution also allows us to address whether the
signature-changing branch can exhibit an effective phantom phase.
Generalized holographic dark-energy models are known to accommodate
both phantom and non-phantom cosmological evolution
\cite{NojiriOdintsov2006,NojiriOdintsov2017}. We do not assume a
particular holographic infrared cutoff here; rather, we characterize
phantom behavior directly through the effective FLRW expansion.
\\
It is important first to distinguish the Einstein and Jordan frames.
In the Einstein frame the scalar field is canonical, and the spatially
flat FLRW equations imply
\[
\dot H_E=-\frac12\dot\psi^{\,2}\leq0 .
\]
Consequently, wherever $H_E\neq0$, the kinematically defined effective
equation-of-state parameter satisfies
\[
w_{\rm eff}^{(E)}
=
-1-\frac{2\dot H_E}{3H_E^2}
=
-1+\frac{\dot\psi^{\,2}}{3H_E^2}
\geq -1 .
\]
Thus no effective phantom regime occurs in the Einstein frame.
\\
The physical Jordan-frame expansion can behave differently because
the gravitational coupling is dynamical. Since in the gauge
$\bar N=1$ the Jordan-frame proper time is
\[
dt_J=F^{-1/2}d\tau ,
\]
we define the physical Jordan-frame Hubble parameter by
\[
\mathcal H_J
:=
\frac{1}{a_J}\frac{da_J}{dt_J}.
\]
Using Eq.~(61), this gives
\[
\mathcal H_J
=
\sqrt{F}
\left(
H_E-\frac12\chi\dot\psi
\right),
\qquad
\chi:=\frac{d\ln F}{d\psi}.
\]
An effective Jordan-frame phantom regime is then characterized
kinematically by
\[
w_{\rm eff}^{(J)}
=
-1-\frac{2}{3\mathcal H_J^2}
\frac{d\mathcal H_J}{dt_J}
<-1,
\]
or equivalently by
\[
\frac{d\mathcal H_J}{dt_J}>0
\]
on an expanding Lorentzian branch.
\\
For the symmetric oscillator solution,
$H_E|_\Sigma=\dot\psi_\Sigma=0$. Moreover, for the positive branch
chosen in Eq.~(83), differentiation of Eq.~(88) yields
\[
\ddot\psi_\Sigma
=
-\sqrt{\frac{8}{3}}\,\Omega_x\Omega_y .
\]
Using $\dot H_E|_\Sigma=0$, one therefore obtains
\[
\left.
\frac{d\mathcal H_J}{dt_J}
\right|_\Sigma
=
-\frac{F_\Sigma}{2}\chi_\Sigma\ddot\psi_\Sigma
=
F_\Sigma\chi_\Sigma
\sqrt{\frac{2}{3}}\,\Omega_x\Omega_y .
\]
Hence, since $F_\Sigma>0$ and $\Omega_x,\Omega_y>0$, the Lorentzian
branch emerging from the transition is locally superaccelerating when
\[
\chi_\Sigma
=
\left(\frac{d\ln F}{d\psi}\right)_\Sigma>0 .
\]
For the explicit realization
$F(\phi)=e^{2q\phi}$ and $\psi=\lambda_0\phi$, one has
\[
\chi=\frac{2q}{\lambda_0},
\qquad
\lambda_0=\sqrt{\omega_0+6q^2}>0,
\]
so the branch considered above exhibits an effective Jordan-frame
phantom regime near the transition for $q>0$.
\\
At the transition hypersurface itself,
$\mathcal H_J|_\Sigma=0$, and therefore
$w_{\rm eff}^{(J)}$ is not defined exactly at $\Sigma$. The statement
above refers to the neighboring expanding Lorentzian region. We also
stress that this is an \emph{effective} phantom behavior generated by
the evolving nonminimal coupling and does not correspond to a
ghostlike scalar degree of freedom: the Einstein-frame scalar retains
a positive canonical kinetic term.
\\
For the critical exponential branch, by contrast, the sign of
$d\mathcal H_J/dt_J$ depends on the nonminimal coupling and the
transition data, so no universal phantom/non-phantom conclusion follows
without further specification of the scalar--tensor model.
\end{remark}

\subsection{Critical exponential realization}

For the critical exponential branch, the Einstein-frame potential is
\begin{equation}
U(\psi)=U_0\exp\left(-\sqrt{\frac{3}{2}}\,\psi\right).
\end{equation}
The corresponding Jordan-frame potential is
\begin{equation}
V(\phi)=F(\phi)^2U(\psi(\phi)).
\end{equation}
Using \(\psi=\lambda_0\phi\), we obtain
\begin{equation}
V(\phi)
=
U_0
\exp\left[
\left(
4q-\sqrt{\frac{3}{2}}\,\lambda_0
\right)\phi
\right].
\label{VexpExplicit}
\end{equation}
Furthermore,
\begin{equation}
\ln F=2q\phi=\frac{2q}{\lambda_0}\psi,
\end{equation}
and hence
\begin{equation}
\chi
:=
\frac{d\ln F}{d\psi}
=
\frac{2q}{\lambda_0}.
\label{chiExplicit}
\end{equation}
Therefore
\begin{equation}
\eta
=
\frac12\chi\sqrt{\frac{2}{3}}
=
\frac{q}{\lambda_0}\sqrt{\frac{2}{3}}.
\label{etaExplicit}
\end{equation}
The Jordan-frame regularity condition for the exponential branch becomes
\begin{equation}
r
=
-\frac{
1+3q\sqrt{2/3}/\lambda_0
}{
1-3q\sqrt{2/3}/\lambda_0
}.
\label{rExplicitExp}
\end{equation}
Thus the nonminimal coupling parameter \(q\) shifts the allowed transition data through the ratio \(r=A_\Sigma/B_\Sigma\). In the minimally coupled limit \(q=0\), one obtains \(r=-1\), recovering the Einstein-frame total-geodesy condition.
\section{Comparison with the Dereli--Tucker and \(f(R,T_\phi)\) constructions}
\label{sec:comparison}

It is useful to compare the present construction with two closely related cases: the original Einstein--scalar Dereli--Tucker model and the trace-coupled \(f(R,T_\phi)\) extension.

First, consider the minimally coupled limit
\begin{equation}
F(\phi)=1,
\qquad
\omega(\phi)=1.
\end{equation}
Then the conformal transformation becomes trivial,
\begin{equation}
\bar g_{\mu\nu}=g_{\mu\nu},
\qquad
\bar a=a,
\qquad
\bar N=N,
\end{equation}
and the Einstein-frame scalar satisfies
\begin{equation}
\psi=\phi,
\end{equation}
up to an irrelevant additive constant. Moreover,
\begin{equation}
V(\phi)=U(\phi).
\end{equation}
In this limit, the oscillator branch reduces directly to the Einstein--scalar Dereli--Tucker construction.

The trace-coupled \(f(R,T_\phi)\) model represents a different type of extension. There, the modification enters through the scalar-field trace \(T_\phi\), and hence changes the effective scalar kinetic and potential sectors in the minisuperspace Lagrangian. Nevertheless, the hyperbolic minisuperspace transformation can still be used, and regular signature-changing solutions survive. In addition, the trace-coupled model admits exact solution sectors beyond the original oscillator branch, including exponential-potential branches.

The scalar--tensor theory studied in the present work differs from both of these cases. The modification is not a matter-trace coupling, but a nonminimal coupling in the gravitational sector,
\begin{equation}
R\longrightarrow F(\phi)R.
\end{equation}
As a result, the effective gravitational coupling becomes dynamical. Although the spatially flat minisuperspace dynamics can be written in Einstein-frame form, the physical Jordan-frame scale factor is
\begin{equation}
a_J=F(\phi)^{-1/2}\bar a.
\end{equation}
Consequently, Einstein-frame regularity does not automatically imply Jordan-frame regularity. The conformal factor must be included in the total-geodesy condition at the signature-changing hypersurface.

The difference between the two branches constructed above can be summarized as follows. In the oscillator branch,
\begin{equation}
H_E\big|_\Sigma=0,
\qquad
\dot\psi_\Sigma=0.
\end{equation}
Therefore, if
\begin{equation}
0<F(\phi_\Sigma)<\infty,
\end{equation}
then the Jordan-frame transition is also totally geodesic:
\begin{equation}
K^{(J)}_{ij}\big|_\Sigma=0.
\end{equation}
In the critical exponential branch, by contrast, the scalar field is generically nonstationary at the transition,
\begin{equation}
\dot\psi_\Sigma\neq0.
\end{equation}
The Jordan-frame total-geodesy condition then requires
\begin{equation}
H_E\big|_\Sigma
=
\frac12
\left(
\frac{d\ln F}{d\psi}
\right)_\Sigma
\dot\psi_\Sigma .
\end{equation}
Thus the dynamical gravitational coupling directly enters the regularity condition.

This is the essential distinction between the scalar--tensor construction and a mere conformal rewriting of the Einstein--scalar solution. In the oscillator branch the conformal factor preserves regularity under mild assumptions. In the exponential branch, however, the nonminimal coupling controls which transition data are admissible in the physical Jordan frame. Moreover, because the scalar is generically nonstationary, the differentiability of the pulled-back Jordan-frame metric must be checked separately if one demands the stronger smooth transverse regularity. The coupling-dependent total-geodesy condition is therefore the genuinely scalar--tensor feature of the present work, while the smoothness of the nonstationary branch is an additional restriction on the allowed scalar--tensor realization.

\section{Conclusion}
\label{sec:conclusion}

In this work we addressed the question of whether a universe can undergo a Euclidean--Lorentzian signature transition when the gravitational coupling itself is dynamical. We studied this problem in scalar--tensor gravity, starting from the Jordan-frame action
\begin{equation}
S=\int d^4x\sqrt{-g}
\left[
\frac12F(\phi)R
-\frac12\omega(\phi)(\nabla\phi)^2
-V(\phi)
\right],
\end{equation}
where the nonminimal coupling \(F(\phi)\) controls the effective gravitational coupling. Since the Jordan frame was taken to be the physical frame, the Einstein frame was used only as a solution-generating representation. The final signature-change and total-geodesy conditions were imposed on the Jordan-frame metric.

After reducing the theory to spatially flat FLRW minisuperspace, we transformed to the Einstein frame and introduced the hyperbolic minisuperspace variables used in the Dereli--Tucker construction. A key technical point was the distinction between the nondegenerate time \(\tau\), in which the minisuperspace equations are solved, and the degenerate signature-changing coordinate \(\beta\), in which the metric takes the form
\begin{equation}
d\bar s^2
=
-\beta\,d\beta^2
+
\bar a(\beta)^2d\vec x^2.
\end{equation}
We used the real pullback prescription
\begin{equation}
\tau-\tau_\Sigma=\frac23\operatorname{sgn}(\beta)|\beta|^{3/2}.
\end{equation}
This prescription makes explicit that total geodesy and smooth transverse regularity are not identical requirements. In particular, a function smooth in \(\tau\) may have only fractional-power regularity as a function of \(\beta\). This distinction is harmless in the stationary branch but becomes important in nonstationary scalar branches.

We constructed two exactly integrable classes of scalar--tensor signature-changing cosmologies. The first is an oscillator--ghost-oscillator branch generalizing the Dereli--Tucker solution. In this branch the Einstein-frame scalar field is stationary at the transition:
\begin{equation}
\dot\psi_\Sigma=0,
\qquad
H_E|_\Sigma=0.
\end{equation}
Consequently, provided the conformal factor is finite, positive, and sufficiently regular at \(\Sigma\), the Jordan-frame transition is regular: the determinant vanishes linearly in the transverse coordinate and the limiting extrinsic curvature of the transition hypersurface vanishes.

The second class is a critical exponential-potential branch. At the critical slope \(\lambda=\sqrt{3/2}\), the light-cone minisuperspace system becomes exactly integrable and the Hamiltonian constraint can be solved in closed form. This branch is qualitatively different from the oscillator branch because the scalar field is generically nonstationary at the transition. As a result, Jordan-frame total geodesy is not automatic. Instead, it imposes the coupling-dependent condition
\begin{equation}
H_E\big|_\Sigma
=
\frac12
\left(
\frac{d\ln F}{d\psi}
\right)_\Sigma
\dot\psi_\Sigma .
\end{equation}
This relation is the main scalar--tensor effect found in this work. It shows that the conformal factor is not a harmless rewriting at a signature-changing hypersurface: the dynamical gravitational coupling directly controls which transition data are admissible in the physical Jordan frame.

The answer to the question posed in the title is therefore affirmative but frame-sensitive. Classical signature change can persist when gravity is dynamical, but its regularity is not simply inherited from the Einstein frame. In the stationary oscillator branch, Jordan-frame regularity follows under mild assumptions on \(F(\phi)\). In the nonstationary exponential branch, the vanishing of the limiting Jordan-frame extrinsic curvature requires a genuine scalar--tensor constraint on the nonminimal coupling. If one demands the stronger Kossowski--Kriele-type smooth transverse regularity, then the pulled-back Jordan-frame metric coefficients must also be checked explicitly. This qualification is essential for interpreting the exponential branch as a fully regular signature-changing geometry.

We have also shown that, although the canonical Einstein-frame scalar
does not admit effective phantom evolution, the dynamical nonminimal
coupling can generate a locally superaccelerating, effective
Jordan-frame phantom regime near the signature-changing surface; for
the explicit oscillator realization considered here this occurs for
$q>0$ on the chosen branch.

The present analysis is restricted to a classical FLRW sector with
fixed spatial topology. Allowing for topology-dependent quantum effects
or dynamical compactification, which can influence the dynamical
selection of spacetime signature
\cite{Elizalde1994,Odintsov1994}, would therefore provide a natural
extension of the present construction.
Other extensions are also worth exploring. Since metric \(f(R)\) gravity
is dynamically equivalent to a scalar--tensor theory with vanishing
Brans--Dicke parameter, the present construction may be applied to
\(f(R)\) models, including Starobinsky-type gravity. It would also be
interesting to investigate whether analogous minisuperspace
transformations exist in teleparallel scalar--torsion gravity and
symmetric teleparallel scalar--nonmetricity gravity, where the FLRW
point-like Lagrangians often have structures close to the
Einstein--scalar system. Finally, the critical exponential branch
suggests a broader question: whether nonstationary scalar transitions
obey general restrictions or no-go results in modified gravity models
admitting Euclidean--Lorentzian signature change.

\end{document}